\documentclass{JHEP3} %{JHEP3}
\JHEPspecialurl{http://jhep.sissa.it/JOURNAL/JHEP3.tar.gz}
\usepackage{amsmath,amssymb,epsfig}

\makeatletter
\newcommand{\fmslash}[2][0mu]{%
  \mathchoice
    {\fmsl@sh\displaystyle{#1}{#2}}%
    {\fmsl@sh\textstyle{#1}{#2}}%
    {\fmsl@sh\scriptstyle{#1}{#2}}%
    {\fmsl@sh\scriptscriptstyle{#1}{#2}}}
\newcommand{\fmsl@sh}[3]{%
  \m@th\ooalign{$\hfil#1\mkern#2/\hfil$\crcr$#1#3$}}

\newcommand{\tr}{\hbox{tr}}

\title{
\begin{center}
On UV/IR mixing in noncommutative\\
 gauge field theories
\end{center}
}

\author{R.Horvat$^1$, A.Ilakovac$^2$, J.Trampetic$^{1,3}$ and J.You$^1$ \\
1. Institute Ruder Bo\v skovi\' c,\\
Bijeni\v{c}ka 54, 10000 Zagreb, Croatia\\
2. Faculty of Science, University of Zagreb\\
Bijeni\v{c}ka 32, 10000 Zagreb, Croatia\\
3. Max-Planck-Institut f\"ur Physik, (Werner-Heisenberg-Institut),
  	 F\"ohringer Ring 6, D-80805 M\"unchen, Germany\\
E-mail: \email{horvat@lei3.irb.hr}, \email{ailakov@rosalind.phy.hr},
\email{josipt@rex.irb.hr}, \email{youjiangyang@gmail.com}}

\abstract{In formulating gauge field theories on noncommutative (NC) spaces
it is suggested that
particles carrying gauge invariant quantities should not be viewed as
pointlike, but rather as extended objects whose sizes grow linearly with
their momenta. This and other generic properties deriving from the nonlocal
character of interactions (showing thus unambiguously their quantum-gravity
origin) lead to a specific form of UV/IR mixing as well as to a pathological
behavior at the quantum level when the noncommutativity parameter $\theta$
is set to be arbitrarily small. In spite of previous suggestions 
that in a NC gauge theory based on the $\theta$-expanded
Seiberg-Witten (SW) maps UV/IR mixing
effects may be under control,
a fairly recent study of photon self-energy
within a  SW $\theta$-exact approach has shown that UV/IR mixing is still
present. We study the self-energy contribution for neutral massless fermions in the
$\theta$-exact approach of NC QED, and show by explicit calculation  that all but one
divergence can be eliminated for a generic choice of the noncommutativity
parameter $\theta$. The remaining  divergence is linked to the pointlike
limit of an extended object.}


\begin{document}

%%%%%%%%%%%%%%%%%%%%%%%%%%%%%%%%%%%%%%%%%%%%%%%%%%%%%%%%%%%%%%%%%%%%%%%%%%%%%%%%%%%%%%%%%

%\section{Introduction}

A reasonable expectation about noncommutative (NC) field theories is that
they should reduce to their commutative relatives whenever the momenta of
the field quanta are well  below $|\theta|^{-1/2}$ (or more exactly when the
limit $\theta \to 0$ is undertaken). In fact that was one of the main
reasons why NC field theories based on star products are so popular. This naive 
expectation is badly violated at the
loop level where the inherent nonlocality of the full theory shows up in the
UV/IR mixing phenomenon \cite{Filk:1996dm,Minwalla:1999px}. 
The effect is characterized by the
appearance of new infrared divergences at the IR-limit of the external
momentum and is accompanied by a nonanalytic behavior in the
noncommutativity parameter $\theta$. In the combined effect, the theory also
shows pathological behavior when the spatial extension of size $|\theta P|$, for
a particle moving with momentum $P$ along the region affected by spacetime
noncommutativity, gets reduced to a point.

Another naive expectation concerns studies of NC gauge theories where one
anticipates the absence of quadratic and linear IR divergences as the
corresponding commutative-theory counterparts deal at most with logarithmic
divergences. That this second expectation does not hold was found for the
first time in \cite{Matusis:2000jf} for
the photon self-energy correction in NC QED, and
reconfirmed later on in \cite{Brandt:2001ud,Brandt:2002if}.
In these papers the $\theta$-dependent
infrared divergent terms (quadratic poles) were found in the real part of
the photon two-point function at one-loop order. It should be however
stressed that such an inappropriate behavior was not found in the imaginary
part of the self-energy, nor in the one-loop correction to the electron
self-energy \cite{Hayakawa:1999yt,Hayakawa:1999zf,Hayakawa:2000zi}, where,
due to miraculous cancellation of phase factors
accompanied with the two vertices, the contribution boils down to the one
found in ordinary QED. We note that this second expectation does hold 
for supersymmetric NC gauge theories, where, because of cancellation between
fermion and boson loops, the UV/IR-mixing problem is softened in such a way
that only logarithmic divergences appear at small values of NC momenta
\cite{Matusis:2000jf}.

The next stage in the development of NC gauge theories occurred with the
seminal paper \cite{Seiberg:1999vs},
where NC fields and gauge transformation parameters
were interpreted as nonlocal, enveloping algebra-valued functions of their
commutative counterparts and of the noncommutativity parameter $\theta$. Such
a connection, known as the Seiberg-Witten map, has virtues compared to the early
attempts based solely on  star products in that now  gauge covariance
\cite{Jackiw:2001jb} and gauge fixing are more easily understood.
This also enables one to
deform commutative gauge theories with essentially arbitrary gauge group and
representation
\cite{Madore:2000en,Jurco:2000ja,Bichl:2001gu,Jurco:2000fb,Jurco:2001my,Jurco:2001rq}.
Such a procedure furthermore  allows the construction of NC
extensions of important particle physic models like the NC standard model
and grand unified theory models
\cite{Calmet:2001na,Behr:2002wx,Aschieri:2002mc,Melic:2005fm,
Melic:2005am,Martin:2011un}. Moreover, upon expanding up to first order in
$\theta$, those models were shown to develop trouble-free one-loop quantum
corrections \cite{Bichl:2001cq,Martin:2002nr,
Brandt:2003fx,Buric:2006wm,Buric:2007qx,Latas:2007eu,
Buric:2007ix,Martin:2009sg,Martin:2009vg,Tamarit:2009iy,Buric:2010wd}.
Still, the Seiberg-Witten map is
also known not to be free of freedom/ambiguities \cite{Asakawa:1999cu,Suo:2001ih}.

It has been argued \cite{Bichl:2001cq} that due to the prodigious freedom
in the Seiberg-Witten map, the photon self-energy in NC QED can be made free from
UV/IR mixing when expansion in the NC parameter $\theta$ is accomplished.
However, it has been
demonstrated recently that in the $\theta$-exact Seiberg-Witten map  UV/IR
mixing yet reappears \cite{Schupp:2008fs}.

A covariant $\theta$-exact approach, inspired by exact formulas for the SW map
\cite{Jurco:2001my,Mehen:2000vs,Okawa:2001mv} and  developed for the 
first time in \cite{Schupp:2008fs}, was used later to study photon-neutrino
phenomenology, namely, photon-neutrino interactions in various
astophysical/cosmological environments \cite{Horvat:2010sr,Horvat:2011iv}.
The central result in
those papers was a derivation of the tree-level coupling of neutrinos with
photons, absent in the standard model settings. There, a gauge field couples
to a spinor field in the adjoint representation of $\rm U_\star(1)$, which enables
particles not charged under the gauge group to have an electric dipole
moment proportional to $\theta$. More generally, electrically neutral matter
fields will be promoted via (hybrid) Seiberg-Witten map to NC fields that
couple via star commutator to photons and transform in the adjoint
representation - this is also the case for phenomenologically promising NC
grand unified theories \cite{Aschieri:2002mc,Martin:2011un}.
Since such an  interaction has no its
commutative counterpart, it switches off when $\theta \to 0$. Such interactions
have already been studied in the framework of NC gauge theories, defined by
the $\theta$-expanded Seiberg-Witten map \cite{Schupp:2002up,Minkowski:2003jg}.
There, like in almost all
other studies of covariant NC field theory an expansion in
$\theta \sim E/\Lambda_{\rm NC}$ was used, where $\Lambda_{\rm NC}$ is
the NC scale and $E$ is the characteristic
energy for a given process. There exists, however, exotic processes like
scattering of ultra high energy neutrinos from extraterrestrial sources,
in which the interacting energy scale runs higher than the current
experimental bound on $\Lambda_{\rm NC}$ \cite{Horvat:2010sr}.
In this case the aforementioned expansion is
no longer applicable, so one is forced to resort to NC field theoretical
frameworks involving the full $\theta$ resummation.

The $\theta$-exact SW map expansion employed in \cite{Schupp:2008fs} could
be derived from
direct recursive computation using consistency conditions
\cite{Horvat:2011iv}. Instead of
expanding in momenta, NC fields are expanded here in powers of the
commutative gauge-field $a_{\mu}$
and hence in the powers of the coupling constant. At each order in $a_{\mu}$,
however, $\theta$-exact expressions can be determined. The motivation behind
this procedure is to sort the interaction vertices by the number of field
operators. In tree-level neutrino-photon coupling, an expansion to the
lowest nontrivial order in $a_{\mu}$ (but all orders in $\theta$) will suffice.

In the present paper, we have undertaken an explicit and exact calculation
of the neutral (massless) fermion self-energy at one loop in NC QED using
the $\theta$-exact Seiberg-Witten map of \cite{Schupp:2008fs}. In spite of pretty
involved and tedious computation, we have managed to express the final result
in an analytic form, so that all pathological behavior originating from
UV/IR mixing can be studied in an unambiguous way \footnote{We stress a
difference with regard to the previous paper \cite{Schupp:2008fs} on this
topic, in which only
a partial structure of the photon self-energy was explicitly evaluated.}. The
structure of the
final result encountered in our calculation is  quite a unique one, not seen
in any of the previous approaches. The crucial  novelty is due to $\theta$
resummation. Namely, although the structure of the self-energy can be guessed by
noting that in addition to the momentum of the neutral fermion, two extra 4-vectors
(${\tilde p}^\mu$ and ${\tilde{\tilde p}}^\mu$,
to be defined below) can be constructed in NC
spacetimes, what we have found here is that the coefficients in front of this new
spinor structures are not constant, but rather functions of the momentum and
of the NC parameter $\theta$ (we give explicit expressions for them). This
brings in consequences for the UV/IR mixing
as well. First of all, the second naive expectation does hold here (without
supersymmetry) as UV/IR mixing is represented by a logarithmic
divergence. We show that if (without loss of generality) $\theta$ is taken to
lie in the $(1,2)$ plane, the  UV/IR mixing term disappears and the rest of
the contribution is well behaved in the UV and IR region, having also a
continuous commutative limit. However, we show that a divergence linked to
the spatial extension of the particle, in a direction of space affected by
$\theta$, still persists.
These are the central results of our paper.

We start with the following model of a Seiberg-Witten type NC
$\rm U_{\star}(1)$ gauge theory:
\begin{equation}
S=\int-\frac{1}{4}F^{\mu\nu}\star F_{\mu\nu}+i\bar\Psi\star\fmslash{D}\Psi\,,
\label{S}
\end{equation}
with definitions of the nonabelian NC covariant derivative and
the field strength, respectively:
\begin{gather}
D_\mu\Psi=\partial_\mu\Psi-i[A_\mu\stackrel{\star}{,}\Psi]\quad\mbox{and}\quad
F_{\mu\nu}=\partial_\mu A_\nu-\partial_\nu
A_\mu-i[A_\mu\stackrel{\star}{,}A_\nu].
\label{DF}
\end{gather}
All the fields in this action are images under (hybrid) Seiberg-Witten maps
of the corresponding commutative fields $a_\mu$ and $\psi$.
In the original work of Seiberg and Witten and in virtually all subsequent 
applications,
these maps are understood as (formal) series in powers of the noncommutativity parameter
$\theta^{\mu\nu}$. Physically, this corresponds to an expansion in momenta and is valid
only for low energy phenomena. Here we shall not subscribe to this point of view and instead
interpret the noncommutative fields as valued in
the enveloping algebra of the underlying gauge group.
This naturally corresponds to an expansion in powers of
the gauge field $a_\mu$ and hence in powers of
the coupling constant $e$. At each order in $a_\mu$ we shall
determine $\theta$-exact expressions.
In the following we discuss the model construction
for the massless fermion case (we set $e=1$ throughout the paper).

In the next step we expand the action in terms of
the commutative gauge parameter $\lambda$
and fields $a_\mu$ and $\psi$
using the following SW map solution \cite{Schupp:2008fs}
\begin{equation}
\begin{split}
A_\mu&=\,a_\mu-\frac{1}{2}\theta^{\nu\rho}{a_\nu}\star_2(\partial_\rho
a_\mu+f_{\rho\mu})+\mathcal O(a^3),
\\
\Psi&=\psi-\theta^{\mu\nu}
{a_\mu}\star_2{\partial_\nu}\psi+\frac{1}{2}\theta^{\mu\nu}\theta^{\rho\sigma}
\bigg\{(a_\rho\star_2(\partial_\sigma
a_\mu+f_{\sigma\mu}))\star_2{\partial_\nu}\psi+2a_\mu{\star_2}
(\partial_\nu(a_\rho{\star_2}\partial_\sigma\psi))\\&-
a_\mu{\star_2}(\partial_\rho
a_\nu{\star_2}\partial_\sigma\psi)-\big(a_\rho\partial_\mu\psi(\partial_\nu
a_\sigma+f_{\nu\sigma})-\partial_\rho\partial_\mu\psi a_\nu
a_\sigma\big)_{\star_3}\bigg\}+\mathcal O(a^3)\psi
\\
\Lambda&=\lambda-\frac{1}{2}\theta^{ij}a_i\star_2\partial_j\lambda+\mathcal
O(a^2)\lambda\,,
\end{split}
\label{SWmap}
\end{equation}
with $\Lambda$ being the NC gauge parameter and
$f_{\mu\nu}$ is the abelian commutative field strength
$f_{\mu\nu}=\partial_\mu a_\nu-\partial_\nu a_\mu$.

The Mojal-Weyl star product $\star$, and its two generalizations,
$\star_2$ and $\star_3$, appearing in (\ref{SWmap}), are defined, respectively, as
\begin{gather}
(f\star g)(x)=f(x)
e^{\frac{i}{2}\overleftarrow{{\partial}_\mu}\,
\theta^{\mu\nu}\,\overrightarrow{{\partial}_\nu}} g(x)\,,
\label{f*g}\\
f(x)\star_2 g(x)=[f(x) \stackrel{\star}{,}g(x)]=\frac{\sin\frac{\partial_1\theta
\partial_2}{2}}{\frac{\partial_1\theta
\partial_2}{2}}f(x_1)g(x_2)\bigg|_{x_1=x_2=x}\,,
\label{f*2g}\\
(f(x)g(x)h(x))_{\star_3}=\left(\frac{\sin(\frac{\partial_2\theta
\partial_3}{2})\sin(\frac{\partial_1\theta(\partial_2+\partial_3)}{2})}
{\frac{(\partial_1+\partial_2)\theta \partial_3}{2}
\frac{\partial_1\theta(\partial_2+\partial_3)}{2}}
+\{1\leftrightarrow 2\}\,\right)f(x_1)g(x_2)h(x_3)\bigg|_{x_i=x}\,,
\label{fgh*3}
\end{gather}
where $\star$-product is associative but noncommutative, while $\star_2$ and $\star_3$
are both commutative but nonassociative. The resulting expansion defines
the one-photon-two-fermion, the two-photon-two-fermion and the three-photon vertices,
$\theta$-exactly.

The expansion of action is straightforward using the SW map expansion \eqref{SWmap}.
In this way, the photon self-interaction terms up to the lowest
nontrivial order are obtained as
\begin{equation}
\begin{split}
S_g=&\int \;i\partial_\mu a_\nu\star[a^\mu\stackrel{\star}{,}a^\nu]
+\frac{1}{2}\partial_\mu
\left(\theta^{\rho\sigma}a_\rho\star_2(\partial_\sigma a_{\nu}+f_{\sigma\nu})\right)\star
f^{\mu\nu}+\mathcal O(a^4).
\end{split}
\label{Sgauge}
\end{equation}
The photon-fermion interaction up to 2-photon-2-fermion terms can derived
by using the first order gauge field and the second order fermion field expansion,
\begin{equation}
\begin{split}
S_f&=\int \;\bar\psi\gamma^\mu[a_\mu\stackrel{\star}{,}\psi]
+(\theta^{ij}\partial_i\bar\psi \star_2
a_j)\gamma^\mu[a_\mu\stackrel{\star}{,}\psi]
%\\&
+i(\theta^{ij}\partial_i\bar\psi
\star_2 a_j)\fmslash\partial\psi-i\bar\psi\star
\fmslash\partial(\theta^{ij}
a_i\star_2\partial_j\psi)\\&
-\!\bar\psi\gamma^\mu[a_\mu\!\stackrel{\star}{,}\!\theta^{ij}
a_i\!\star_2\!\partial_j\psi]\!-\!\bar\psi\gamma^\mu
[\frac{1}{2}\theta^{ij}a_i\!\star_2\!(\partial_j
a_\mu\!+\!f_{j\mu})\!\stackrel{\star}{,}\!\psi]\!
-\!i(\theta^{ij}\partial_i\bar\psi
\!\star_2\!a_j)\fmslash\partial(\theta^{kl}
a_k\!\star_2\!\partial_l\psi)\\&+\frac{i}{2}\theta^{ij}\theta^{kl}
\big[(a_k\star_2(\partial_l
a_i+f_{li}))\star_2\partial_j\bar\psi
+2a_i\star_2(\partial_j(a_k\star_2\partial_l\bar\psi))-
a_i\star_2(\partial_k
a_j\star_2\partial_l\bar\psi)\\&+\big(a_i\partial_k\bar\psi(\partial_j
a_l+f_{jl})-\partial_k\partial_i\bar\psi a_j a_l\big)_{\star_3}\big]
\fmslash\partial\psi+\frac{i}{2}\theta^{ij}\theta^{kl}\bar\psi\fmslash\partial
\big[(a_k\star_2(\partial_l
a_i+f_{li}))\star_2\partial_j\psi\\&+\!2a_i\!\star_2\!
(\partial_j(a_k\!\star_2\!\partial_l\psi))\!-\!a_i\!\star_2\!(\partial_k
a_j\!\star_2\!\partial_l\psi)\!+\!
\big(a_i\partial_k\psi(\partial_j
a_l\!+\!f_{jl})\!-\!\partial_k\partial_i\psi a_j
a_l\big)_{\star_3}\big]\\&+\bar\psi\mathcal{O}(a^3)\psi.
\end{split}
\label{Sfermion}
\end{equation}
Note that actions for gauge and matter fields obtained above,
(\ref{Sgauge}) and (\ref{Sfermion}) respectively, are nonlocal
objects due to the presence of the (generalized) 
star products: $\star$, $\star_2$ and $\star_3$.

\begin{figure}
\begin{center}
\includegraphics[width=70mm]{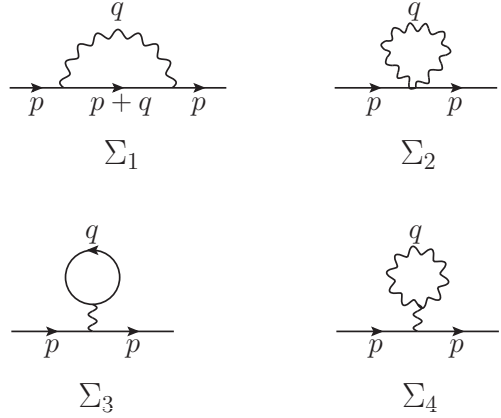}
\end{center}
\caption{One-loop self-energy of a neutral (massless) fermion}
\label{Sigma1-loop}
\end{figure}

As depicted in Fig. \ref{Sigma1-loop}, there are four Feynman diagrams
contributing  to the neutral massless fermion self-energy at one-loop: 
the bubble diagram ($\Sigma_1$),
the 3-fields tadpole with fermion/photon loop 
($\Sigma_3$ and $\Sigma_4$, respectively), and the
fourth one is the 4-field (2-fermion-2-photon) tadpole ($\Sigma_2$).
Only the $\Sigma_1$ and $\Sigma_3$ contributions have their commutative-theory analogs. With
the aid of (\ref{Sfermion}), we have verified by explicit calculation that the 4-field
tadpole ($\Sigma_2$) does vanish. The 3-fields tadpoles ($\Sigma_3$ 
and $\Sigma_4$) can be ruled out by invoking
the NC charge conjugation symmetry \cite{Aschieri:2002mc}
\footnote{Here we take the charge conjugation transformation to be the same
as equation (64) to (66) in \cite{Aschieri:2002mc},
i.e. ${\theta^C}^{\mu\nu}=-\theta^{\mu\nu}$.},
so we do not take them into account here.
Thus only the bubble diagram needs to be evaluated. By extracting the
relevant Feynman rule from (\ref{Sfermion}), one obtains in spacetime of the
dimensionality $D$,
\begin{eqnarray}
%\begin{split}
\Sigma_1
&=&\mu^{4-D}\int \frac{d^D q}{(2\pi)^D}
\bigg(\frac{\sin\frac{q\theta p}{2}}{\frac{q\theta
p}{2}}\bigg)^2\frac{1}{q^2}\frac{1}{(p+q)^2}\frac{1\pm\gamma^5}{2}
\nonumber\\
&\cdot & \{(q\theta p)^2(4-D)(\fmslash p+\fmslash q)
\label{Sigma1}\\
&+&(q\theta p)\left[/\!\!\tilde q(2p^2+2p\cdot q)-/\!\!\tilde
p(2q^2+2p\cdot q)\right]
\nonumber\\
&+&\left[\fmslash p(\tilde q^2(p^2+2p\cdot q)-q^2(\tilde p^2+2\tilde
p\cdot\tilde q) )+\fmslash q(\tilde p^2(q^2+2p\cdot q)-p^2(\tilde
q^2+2\tilde p\cdot\tilde q ))\right]\}\,,
\nonumber
%\end{split}
\end{eqnarray}
where $({\tilde p}^\mu=(\theta p)^\mu=\theta^{\mu\nu} p_\nu )$,
and in addition $({\tilde{\tilde p}}^\mu=(\theta\theta p)^\mu
=(\theta^{\mu\nu}\theta_{\nu\rho}p^\rho)$.

To perform computations of those integrals using the dimensional
regularization method, we first use
the Feynman parametrization on the quadratic denominators,
then the Heavy Quark Effective theory (HQET)
parametrization \cite{Grozin:2000cm} is used to combine
the quadratic and linear denominators.
In the next stage we use the Schwinger
parametrization to turn the denominators
into Gaussian integrals. The outcome is
then combined with different phase factors
emerging from the sine term in (\ref{Sigma1}).

Evaluating the relevant integrals for
$D=4-\epsilon$ in the limit $\epsilon\to 0$, we obtain the final
expression for the self-energy as
\begin{equation}
\Sigma_1=\gamma_{\mu}
\bigg[p^{\mu}\: A+(\theta{\theta p})^{\mu}\;\frac{p^2}{(\theta p)^2}\;B\bigg]\,,
\label{sigma1AB}
\end{equation}
where
\begin{eqnarray}
&\hspace{-.6cm}A&= \frac{-1}{(4\pi)^2}
\bigg[p^2\;\bigg(\frac{\tr\theta\theta}{(\theta p)^2}
+2\frac{(\theta\theta p)^2}{(\theta p)^4}\bigg) A_1 
+ \bigg(1+p^2\;\bigg(\frac{\tr\theta\theta}{(\theta p)^2}
+\frac{(\theta\theta p)^2}{(\theta p)^4}\bigg)\bigg)A_2 \Bigg]\,,
\label{A}\\
&\hspace{-.6cm}A_1&=\frac{2}{\epsilon}+\ln(\mu^2(\theta p)^2)
+ \ln({\pi e^{\gamma_{\rm E}}})
%\nonumber\\
%&+&
+\sum\limits_{k=1}^\infty \frac{\left(p^2(\theta p)^2/4\right)^k}{\Gamma(2k+2)}
 %\left(\frac{p^2(\theta p)^2)^k}{4}\right)
\left(\ln\frac{p^2(\theta p)^2}{4} + 2\psi_0(2k+2)\right)
 \,,
\label{A1}\\
&\hspace{-.6cm}A_2&=-\frac{(4\pi)^2}{2} B = -2
\nonumber\\
&\hspace{-.6cm}+& \sum\limits_{k=0}^\infty
\frac{\left(p^2(\theta p)^2/4\right)^{k+1}}{(2k+1)(2k+3)\Gamma(2k+2)}
%\left(\frac{(p^2(\theta p)^2)^{k+1}}{4}\right)
\left(\ln\frac{p^2(\theta p)^2}{4} - 2\psi_0(2k+2)
- \frac{8(k+1)}{(2k+1)(2k+3)} \right),
\label{A2}
\end{eqnarray}
with  $\gamma_{\rm E}\simeq0.577216$ being Euler's constant.
It is to be noted here that the spinor structure proportional to
$\tilde{\fmslash p}$ is missing in the final result. This conforms with the
calculation of the neutral fermion self-energy in the $\theta$-expanded SW map approach 
\cite{Ettefaghi:2007zz}.

The $1/\epsilon$ UV divergence
could in principle be removed by a properly chosen counterterm.
However (as already mentioned) due to the specific momentum-dependent 
coefficient in front of it, a nonlocal 
form for it is required. It is important to stress here
that amongst other
terms contained in both coefficients $A_1$ and $A_2$, there are structures
proportional to
\begin{equation}
\left({p^2(\theta p)^2}\right)^{n+1}(\ln{(p^2(\theta p)^2)})^m,\;\;
{\forall n} \;\; {\rm and} \;\;m=0,1.
\label{lnUV/IR'}
\end{equation}
The numerical factors in front of the above structures are rapidly-decaying,
thus series are always convergent for finite argument, 
as we demonstrate in \cite{arXiv:1111.4951}.

Turning to the UV/IR mixing problem, we recognize
a soft UV/IR mixing term represented by a logarithm,
\begin{equation}
\Sigma_{\rm UV/IR}=-{\fmslash p}\,p^2 \,
\bigg(\frac{\tr\theta\theta}{(\theta p)^2}
+2\frac{(\theta\theta p)^2}{(\theta p)^4}\bigg)\cdot
\frac{2}{(4\pi)^2}\ln{|\mu(\theta p)|}.
\label{lnUV/IR}
\end{equation}
Thus, we see that the second naive expectation about NC gauge field theories
does hold here even without invoking supersymmetry.

Instead of dealing with nonlocal counterterms, we take a different route
here to  cope with various divergences besetting (\ref{sigma1AB}). Since $\theta^{0i}
\neq 0$ makes a NC theory nonunitary \cite{Gomis:2000zz},
we can, without loss of generality,
chose $\theta$ to lie in the (1, 2) plane
\begin{equation}
\theta^{\mu\nu}=\frac{1}{\Lambda_{\rm NC}^2}
\begin{pmatrix}
0&0&0&0\\
0&0&1&0\\
0&-1&0&0\\
0&0&0&0
\end{pmatrix}.
\label{degen}
\end{equation}
Automatically, this produces
\begin{equation}
\frac{\tr\theta\theta}{(\theta p)^2}
+2\frac{(\theta\theta p)^2}{(\theta p)^4}=0,\:\forall p.
\label{degen1}
\end{equation}

With (\ref{degen1}), $\Sigma_1$,
in terms of Euclidean momenta, receives the following form:
\begin{equation}
\Sigma_1=\frac{-1}{(4\pi)^2}\gamma_\mu\left[p^\mu
\bigg(1 + \frac{\tr\theta\theta}{2}\frac{p^2}{(\theta p)^2}\bigg)
-2(\theta\theta p)^\mu\frac{p^2}{(\theta p)^2} \right] A_2.
\label{Sigmabuble}
\end{equation}
By inspecting (\ref{A2}) one can be easily convinced that $A_2$ is 
free from the $1/\epsilon$ divergence and the UV/IR mixing term, being also 
well-behaved in the infrared, in the $\theta \rightarrow 0$ 
as well as $\theta p \rightarrow 0$
limit. We see, however, that the two terms in (\ref{Sigmabuble}), 
one being proportional to
$\fmslash{p}$ and the other proportional to $\fmslash{\tilde{\tilde p}}$,
are still ill-behaved in the $\theta p \rightarrow 0$
limit. If, for the choice (\ref{degen}), $P$ denotes the momentum in the (1, 2)
plane, then $\theta p = \theta P$. For instance, a particle moving
inside the noncommutative plane with
momentum $P$ along the one axis, has a spatial extension of size $|\theta P|$
along the other. For the choice (\ref{degen}), $\theta p \rightarrow 0$ corresponds to
a zero momentum projection onto the (1, 2) plane. Thus, albeit in our approach the
commutative limit ($\theta \rightarrow 0$) is smooth at the quantum level,
the limit when an extended object (arising due to the fuzziness of space)
shrinks to zero, is not. We could surely claim that in our approach the
UV/IR mixing problem is considerably softened; on the other hand, we have
witnessed how the problem strikes back in an unexpected way. This is, at the
same time, the first example where this two limits are not degenerate.

Summing up, we have demonstrated how quantum effects in the $\theta$-exact
Seiberg-Witten map approach to NC gauge field theory  reveal a much richer
structure for the one-loop quantum correction to the fermion two point
function (and accordingly for the UV/IR mixing problem), than 
observed previously in
approximate models restricting to low-energy phenomena. In our approach, 
UV/IR mixing assumes a new form where the commutative limit and the limit of
zero size of the extended object are fully disentangled. Our analysis  can be
considered trustworthy since we
have obtained the final result in an analytic, closed-form manner.
We believe that a promising avenue of research would be
to use the enormous freedom in the
Seiberg-Witten map to look for other forms which UV/IR mixing may assume.
One alternative form has been already found \cite{arXiv:1111.4951}. 
Finally, we mention that our
approach to UV/IR mixing should not be confused with the one  based on
a theory with UV completion ($\Lambda_{\rm UV} < \infty$), where a theory
becomes an effective QFT, and the UV/IR mixing manifests itself via a
specific relationship between
the UV and the IR cutoffs
\cite{AlvarezGaume:2003mb,Jaeckel:2005wt,Abel:2006wj,Horvat:2010km}.

%\section{One point function and charge conjugation symmetry}\label{discrete}
\vspace{.5cm}
%\section{Acknowledgment}
We thank to D. Bahns for many useful discussions and to P. Schupp
for critically reading the manuscript. 
J.T. would like to acknowledge support of 
Alexander von Humboldt Foundation
(KRO 1028995), Max-Planck-Institute for Physics, Munich, for hospitality, 
and W. Hollik for fruitful discussions.
The work of R.H. and J.T. are supported by
the Croatian Ministry of Science, Education and Sports
under Contracts Nos. 0098-0982930-2872 and 0098-0982930-2900, respectively.
The work of A.I. supported by
the Croatian Ministry of Science, Education and Sports
under Contracts Nos. 0098-0982930-1016.
The work of J.Y. was supported by the Croatian NSF and the IRB Zagreb.
%and by the German Research Foundation (Deutsche
%Forschungsgemeinschaft (DFG)) through the Institutional Strategy of the
%University of G\"ottingen.

%\bibliography{thesis}
%\bibliographystyle{JHEP}

\end{document}